\documentclass[12pt,english]{article}
\usepackage[T1]{fontenc}
\usepackage[latin9]{inputenc}
\usepackage{geometry}
\usepackage{babel}
\usepackage{float}
\usepackage{url}
\usepackage{amsmath}
\usepackage{amssymb}
\usepackage{graphicx}
\usepackage{esint}
\usepackage[unicode=true,
 bookmarks=false,
 breaklinks=true,pdfborder={0 0 1},backref=section,colorlinks=false]
 {hyperref}

\makeatletter
\@ifundefined{date}{}{\date{}}

\usepackage{babel}
\usepackage{url}

\usepackage{babel}

\newtheorem{OurClaim}{Claim}

\makeatother

\begin{document}

\title{The deceiving simplicity of problems with infinite charge distributions
in electrostatics}

\author{Marcin Ko\'{s}cielecki$^{1}$, Piotr Nie\.{z}urawski$^{2}$ \\
 {\small{}$^{1}$Department of Mathematical Methods in Physics, Faculty
of Physics, University~of~Warsaw } \\
 {\small{}ul.~Pasteura~5, 02-093 Warsaw, Poland } \\
 {\small{}$^{2}$Institute of Experimental Physics, Faculty of Physics,
University of Warsaw } \\
 {\small{}ul.~Pasteura~5, 02-093 Warsaw, Poland } \\
 \textit{\small{}marcin.koscielecki@fuw.edu.pl}{\small{}, }\textit{\small{}piotr.niezurawski@fuw.edu.pl}{\small{}
} }

\maketitle
\abstract{We show that for an infinite, uniformly charged plate
no well defined electric field exists in the framework of electrostatics,
because it cannot be defined as a mathematically consistent limit
of a solution for a finite plate. We discuss an infinite wire and
an infinite stripe as examples of infinite charge distributions for
which the electric field can be determined as a limit in a formal,
mathematical way. We also propose a didactic framework that can help
students understand subtleties related to the problems of limits in
electrostatics. The framework consists of heuristic tools (claims)
that help to align intuitions in the spirit of a rigorous definition
of an integral. We thoroughly discuss to what degree the solution
for a finite plate agrees with the traditional but unfortunately ill-defined
solution for an infinite plate. Physics is a science of approximations.
One can ask why the use of mathematically ill-defined formulae and
objects should be forbidden if they make life simpler. In our opinion,
approximations should have solid physical and mathematical foundations.

}

\section{Introduction}

In this paper, we discuss conceptual problems related to teaching
electrostatics to college students. Many exercises involve sophisticated
integrating over bounded or unbounded domains. However, the problem
of the existence of integrals over unbounded domains is rarely discussed.
Generally, the teaching process focuses on the application of \textquotedbl{}symmetry\textquotedbl{}
as a leading heuristic rule, but a mathematical perspective on validity
and the drawbacks of such an approach are not presented, even in standard
textbooks (e.g. \cite{Dobbs}, \cite{Feynman2013}, \cite{Griffiths2013electrodynamics},
\cite{halliday2014fundamentals}, \cite{halliday2014instructors},
\cite{herbert1991introductory}, \cite{MIT}, \cite{Prytz2015}).
The absence of such a discussion is permanent and hard to accept.\textbf{
}Students who attend lectures have completed at least a basic calculus
course and should be capable of understanding explanations related
to the existence of limits, the Riemann integral over an unbounded
domain and the integral in the Cauchy principal value sense. More
than fifty years ago R.~Shaw \cite{Shaw-Symmetry-Uniq} expressed
his frustration in the following words: 
\begin{quote}
\emph{Presumably not unconnected with this uncritical acceptance of
arguments based on symmetry is the fact that false, or at best incomplete,
arguments of this type are quite common in elementary textbooks on
electricity.} 
\end{quote}
We will show that the \textquotedbl{}symmetry heuristics\textquotedbl{}
in electrostatics do more harm than good and do not agree with the
formal mathematical definition of limit. Even the Cauchy principal
value, sometimes presented as a mathematical representation of a \textquotedbl{}symmetry
heuristics\textquotedbl{}, does not work in the long run as it clashes
with invariance under translations. We understand that a heuristic
is necessary to frame student intuition and give a general feeling
of the subject \cite{Sherin-How-students}.\textbf{ }Attempts to ``associate
meaning with certain structures'' in case of definite integral in
the context of electrostatics are presented in \cite{Meredith-Context,Doughty-cues}.
However, we did not find any discussion about a ``concept image''
related to integration over an unbounded domain. Therefore, we propose
a new leading concept for the case of charge distributions over unbounded
domains.

Unbounded distributions are problematic in various aspects. Here we
focus on the existence of electric field integrals. However, other
approaches are present in the literature. For example, the authors
of \cite{Palma} discuss asymptotic conditions of an unbounded charge
distribution necessary to obtain the assumed asymptotics of the potential.
We show our ideas in action discussing a few examples of unbounded
charge distributions: the infinite wire, the infinite stripe, a quarter
of the infinite plate, and the infinite plate.

We disagree with the popular opinion that calculating the electric
field of the infinite plate is the simplest and correct way to obtain
the approximation of the field of a big but finite plate. Let us assume
that we somehow convince a student that for a large plate, far from
its edges the field should be \emph{nearly} uniform and \emph{nearly}
perpendicular to the plate. The student uses textbook procedures and
receives the result. This approach has three significant flaws. First,
the student has no idea how precise is the result. What is the error
of the result? Is it $10\%$ or $10^{-6}\%$? (for a detailed discussion
see chapter \ref{sec:How-important-are}) Second, this approach strengthens
the conviction that the field of the infinite plate exists as -- intuitively
but not mathematically -- the limit of the \emph{enlarging} procedure.
Third, from the beginning the student is exposed to dirty tricks dressed
up as fundamental principles.


\section{Integrals over unbounded domains in electrostatics}

\subsection{The didactic challenge}

The electric field of uniformly charged infinite objects such as an
infinite wire and a plate is one of the standard topics present in
introductory courses in electrostatics. Given some specific volumetric
distribution of charges $\rho(\vec{r})$ confined in some finite volume
(domain) $V\in\mathbb{R}^{3}$, the electric field at point $\vec{r}$
is given by the formula: 
\begin{equation}
\vec{E}(\vec{r})=k\int_{V}\frac{\rho(\vec{r}')(\vec{r}-\vec{r}')}{|\vec{r}-\vec{r}'|^{3}}\mbox{d}V'\label{r1}
\end{equation}
where $k=\frac{1}{4\pi\varepsilon_{0}}$. In the case of infinite
volume, the integral of the electric field over a non-compact domain
should be computed as a limit: 
\begin{equation}
\vec{E}_{\infty}(\vec{r})=\lim_{V\rightarrow{\mathbb{R}}^{3}}k\int_{V}\frac{\rho(\vec{r}')(\vec{r}-\vec{r}')}{|\vec{r}-\vec{r}'|^{3}}\mbox{d}V'\label{r2-limit-to-infinity}
\end{equation}
The existence of the limit (\ref{r2-limit-to-infinity}) is treated
as a default in textbooks. Authors of textbooks (e.g. \cite{Dobbs},
\cite{Feynman2013}, \cite{Griffiths2013electrodynamics}, \cite{halliday2014fundamentals},
\cite{halliday2014instructors}, \cite{herbert1991introductory},
\cite{MIT}, \cite{Prytz2015}) 
 implicitly assume that integrals over unbounded domains are computable
and they focus on presenting the most effective ways to calculate
the limit (\ref{r2-limit-to-infinity}) (often using Gauss's law),
so the discussion has a technical and not an existential nature --
for more details see Appendix \ref{sec:Examples-of-inconsistencies}.
Unfortunately, a discussion about the existence of limit (\ref{r2-limit-to-infinity})
is unavoidable, even in the case of such a standard problem of electrostatics
as a charged infinite plate. The absence of such a discussion is difficult
to understand. One of the pessimistic explanations can be found in
\cite{Bohren2009}. However, we optimistically believe that the authors
could not find a satisfying way to explain all the subtleties to students.
Indeed, comments like \cite{Jeffreys} (p.~181) do not help: 
\begin{quote}
\label{test} A double integral $\int\! f(x,y)\,\mbox{d}x\,\mbox{d}y$
over an infinite region $R$ can be defined by taking a sequence of
regions $\{R_{n}\}$ such that, for any part of $R$, this part is
included in all $R_{n}$ for $n$ greater than some $m$. If the double
integral over $R_{n}$ has a unique limit for all such sequences,
this limit can be taken as the definition of the integral over $R$.
Improper double integrals may be defined similarly. It appears, however,
that unless the same process gives a unique value when $|f(x,y)|$
is substituted for $f(x,y)$ the value of the limit will depend on
the shapes of the regions $R_{n}$, and consequently a non-absolutely
convergent double integral has no meaning unless these are specified. 
\end{quote}
However true, these thoughts are convoluted enough to present a didactic
challenge. Unfortunately, the over-abundant \textquotedbl{}symmetry
heuristics\textquotedbl{} presented as obvious in textbooks makes
detailed discussion about the existence of limit $(\ref{r2-limit-to-infinity})$
more difficult. The didactic challenge is solvable but to do this
the \textquotedbl{}symmetry\textquotedbl{} argument should not be
used as a leading idea in electrostatics. A concise presentation of
problems related to limit (\ref{r2-limit-to-infinity}) could involve
the following steps: 
\begin{enumerate}
\item Downgrade \textquotedbl{}symmetry\textquotedbl{} intuitions as they
do not help with the nuances of calculations over unbounded domains. 
\item Find intuitions/heuristics that help to understand the mathematical
subtleties of limit (\ref{r2-limit-to-infinity}). 
\item Check which classical problems of electrostatics can be computed directly
from definition (\ref{r2-limit-to-infinity}). 
\item Accept the fact that some problems become ill-posed when extended
to an unbounded domain. 
\item Discuss the finite domain solutions for non-extendable problems. 
\end{enumerate}

\subsection{Drawbacks of symmetry intuitions}

We present simple examples of how intuition built on the \textquotedbl{}symmetry\textquotedbl{}
argument conflicts with strict mathematical definitions. We believe
that the typical second year student is capable of understanding the
examples that follow.

\subsubsection{Limits}

To show that the limit 
\begin{equation}
\lim_{x\rightarrow+\infty}\cos(x)\label{sin}
\end{equation}
does not exist (see also: \cite{Bourchtein}, p.~66), it is enough
to show a counterexample -- for two different sequences: $x_{n}=2\pi n$,
and $y_{n}=\pi+2\pi n$, $n\in\mathbb{N}$, the limit (\ref{sin})
gives two different results: 
\begin{equation}
\lim_{n\rightarrow+\infty}\cos\left(2\pi n\right)=1,\quad\lim_{n\rightarrow+\infty}\cos\left(\pi+2\pi n\right)=-1.\label{sin1}
\end{equation}
One would get into serious trouble during a calculus exam arguing
that 
\begin{equation}
\lim_{x\rightarrow+\infty}\cos(x)=0,\label{sin_z}
\end{equation}
using the \textquotedbl{}let's take the average\textquotedbl{} or
\textquotedbl{}symmetry with respect to the $x$-axis\textquotedbl{}
argument, even if 
\begin{equation}
\lim_{n\rightarrow\infty}\cos\left(\frac{\pi}{2}+\pi n\right)=0.\label{sin_zero}
\end{equation}
The truth is that not every sequence has a limit.

\subsubsection{Integrals}

Imagine one has to compute the integral of a real function $f(x)$
over $\mathbb{R}$. The existence of such an integral, by definition,
is related to the existence of two independent limits: 
\begin{equation}
\int_{-\infty}^{+\infty}f(x)\,\mbox{d}x:=\lim_{A\rightarrow-\infty}\lim_{B\rightarrow+\infty}\int_{A}^{B}f(x)\,\mbox{d}x\label{int}
\end{equation}
In this spirit the integral: 
\begin{equation}
\int_{-\infty}^{+\infty}\sin(x)\,\mbox{d}x=\lim_{A\rightarrow-\infty}\lim_{B\rightarrow+\infty}\int_{A}^{B}\sin(x)\,\mbox{d}x=\lim_{B\rightarrow+\infty}(-\cos(B))-\lim_{A\rightarrow-\infty}(-\cos(A))\label{int_sin_minus_inf_to_inf}
\end{equation}
does not exist because each limit for $\cos(x)$ does not exist as
we have shown in (\ref{sin1}).

Why cannot one use the argument of symmetry and claim that the integral
(\ref{int_sin_minus_inf_to_inf}) is equal to zero because the $\sin(x)$
is an odd function? The symmetry argument applied here essentially
means that we treat variables $A$ and $B$ as not independent --
now we impose an additional constraint $A=-B$ and we want to solve
problem (\ref{int_sin_minus_inf_to_inf}) by computing: 
\begin{equation}
\lim_{B\rightarrow+\infty}\int_{-B}^{B}\sin(x)\,\mbox{d}x=\lim_{B\rightarrow+\infty}(-\cos(B))-(-\cos(-B))=0,\label{int_sin_minus_B_to_B}
\end{equation}
However from a mathematical point of view, the value of integral (\ref{int_sin_minus_inf_to_inf})
is equal to (\ref{int_sin_minus_B_to_B}) only when (\ref{int_sin_minus_inf_to_inf})
exists in the sense of (\ref{int})! The nuance lies in the implication:
if the integral defined in (\ref{int}) exists then the result does
not depend on the way we link the $A$ and $B$ values, say $A=-B^{2}$
or $A=-2B$, etc. But the converse is not true.

To save the \textquotedbl{}symmetry\textquotedbl{} argument one could
abandon formal definitions and say that every integral in electrostatics
should be understood in the \textquotedbl{}symmetric\textquotedbl{}
sense: 
\begin{equation}
\operatorname{v.p.}\int_{-\infty}^{+\infty}f(x)\,\mbox{d}x:=\lim_{A\rightarrow\infty}\int_{-A}^{A}f(x)\,\mbox{d}x\label{intvp}
\end{equation}
where $\operatorname{v.p.}$ means the Cauchy principal value (see
also: \cite{Gelbaum}, p.~45, example $11$). Unfortunately such
an approach also clashes with the \textquotedbl{}symmetry\textquotedbl{}
heuristic when one tries to apply it to symmetric functions such as
$\cos(x)$: 
\begin{equation}
\lim_{A\rightarrow+\infty}\int_{-A}^{A}\cos(x)\,\mbox{d}x=\lim_{A\rightarrow+\infty}\sin(A)-\sin(-A)=\lim_{A\rightarrow+\infty}2\sin(A)\label{AAA}
\end{equation}
It is easy to show, as we did for (\ref{sin}), that the last limit
in ($\ref{AAA}$) does not exist. The conflict also manifests itself
at the level of intuitions. Physicists like the idea of translational
invariance as much as symmetry. On the computational level this means
that the integral (in the principal value sense as well) over an unbounded
domain should not change if we shift the graph of the function by
$\frac{\pi}{2}$, so the result for $\sin(x)$ should be the same
as for $\cos(x)$.

\section{A conceptual framework for understanding electrostatics}

We would like our students posess the ability to first think about
whether a problem has a solution before going into the technical nuances
of finding the best shortcut for solving it. The first step should
not involve a discussion about the possible symmetries of the problem
for it treats the existence of solutions by default. We need a leading
idea that focuses on the nuances of the existence of limit (\ref{r2-limit-to-infinity})
and at the same time could be accepted on the heuristic level as it
relates to physical objects. Therefore we propose two equivalent claims:

\begin{OurClaim}\label{claim-1} The property of a system should
not depend on the method of dividing the system into subsystems. \end{OurClaim}

\begin{OurClaim}\label{claim-2} The property of the system should
not depend on the method of constructing the system from subsystems.
\end{OurClaim}

The above claims consider two important facts related to limit (\ref{r2-limit-to-infinity}):
1) The existence of limit for an unbounded region means that all possible
ways to fill--up that region must lead to the same result. 2) If the
result of (\ref{r2-limit-to-infinity}) is not independent of the
choice of division into smaller parts, the limit does not exist. Claim
\ref{claim-1} represents a static approach to the system while Claim
\ref{claim-2} focuses on its dynamical aspect. Both should appeal
to different mathematical and physical intuitions of our students.

In light of the above claims students would be less surprised to see
that the integral (\ref{r2-limit-to-infinity}) in the case of an
infinite charged plate gives different values depending on the particular
prescription of extending volume $V$ (in a two-dimensional case)
to infinity. Students can check that such a field, understood as a
unique solution of (\ref{r2-limit-to-infinity}), does not exist and
has the same meaningless status as limit (\ref{sin}). In the next
sections we will revisit standard problems of electrostatics and use
Claims \ref{claim-1} and \ref{claim-2} as the leading ideas.

\section{Classical problems of electrostatics revisited\label{sec:Discussion-of-infinite}}

\subsection{The didactic challenge, part II}

We aim to show that the application of Claims \ref{claim-1} and \ref{claim-2}
can lead to interesting results or can at least provoke refreshing
discussions with students. We examine the existence of the electric
field for: a uniformly charged infinite wire, an infinite stripe and
an infinite plate by computing appropriate limits of solutions for
a finite wire and a rectangle. For linear and surface charge distributions,
we use the following variants of formula (\ref{r1})

\begin{equation}
\vec{E}(\vec{r})=k\int_{L}\frac{\lambda(\vec{r}')(\vec{r}-\vec{r}')}{|\vec{r}-\vec{r}'|^{3}}\mbox{d}L'\label{r1-linear}
\end{equation}

where $\lambda(\vec{r})$ is a linear charge distribution along some
finite length curve $L$, and

\begin{equation}
\vec{E}(\vec{r})=k\int_{S}\frac{\sigma(\vec{r}')(\vec{r}-\vec{r}')}{|\vec{r}-\vec{r}'|^{3}}\mbox{d}S'\label{r1-surface}
\end{equation}

where $\sigma(\vec{r})$ is a surface charge distribution on some
finite area surface $S$. We do not discuss how to derive (\ref{r1-linear})
and (\ref{r1-surface}) from (\ref{r1}) by treating charge density
in the rigorous, distributive sense. Such an approach, however preferable,
would pose another didactic challenge as first year students are not
familiar with the theory of distributions.

\subsection{From finite to infinite straight wire\label{sub:From-finite-to-Wire}}

First we consider a one-dimensional, straight, uniformly charged wire
with linear charge density $\lambda$. We start with a wire $L$ of
finite length extending from point $a$ to $b$ on the $X$ axis.
We determine the electric field at point $\vec{r}=[0,\, y,\, z]$,
assuming $y\neq0$ or $z\neq0$. Using Coulomb's law and superposing
contributions from infinitesimal charge elements $\lambda\,\mbox{d}x'$
at point $\vec{r}'=[x',\,0,\,0]$ one obtains:

\begin{eqnarray*}
\vec{E}(\vec{r}) & = & k\lambda\int_{L}\frac{\vec{r}-\vec{r}'}{|\vec{r}-\vec{r}'|^{3}}\mbox{d}x'
\end{eqnarray*}

where

\[
\int_{L}\frac{\vec{r}-\vec{r}'}{|\vec{r}-\vec{r}'|^{3}}\mbox{d}x'=\hat{e}_{x}\int_{L}\frac{-x'}{|\vec{r}-\vec{r}'|^{3}}\mbox{d}x'+\hat{e}_{y}\int_{L}\frac{y}{|\vec{r}-\vec{r}'|^{3}}\mbox{d}x'+\hat{e}_{z}\int_{L}\frac{z}{|\vec{r}-\vec{r}'|^{3}}\mbox{d}x'
\]

and

\begin{eqnarray*}
\vec{r}-\vec{r}' & = & [-x',\, y,\, z]\\
|\vec{r}-\vec{r}'| & = & \sqrt{x'^{2}+y^{2}+z^{2}}
\end{eqnarray*}

As the $y$ component of the electric field is analogous to the $z$
component, for simplicity we continue calculation of the field at
point $\vec{r}=[0,\,0,\, z]$, on the $Z$ axis (assuming $z\neq0$).
In this case $E_{y}=0$. We calculate $x$ and $z$ components of
$\vec{E}$:

\begin{align}
E_{x} & =k\lambda\int_{a}^{b}\frac{-x'}{\sqrt{x'^{2}+z^{2}}^{3}}\mbox{d}x'=k\lambda\left(\frac{1}{\sqrt{b^{2}+z^{2}}}-\frac{1}{\sqrt{a^{2}+z^{2}}}\right)\label{eq:wire-ExEz-integrals}\\
E_{z} & =k\lambda\int_{a}^{b}\frac{z}{\sqrt{x'^{2}+z^{2}}^{3}}\mbox{d}x'=k\lambda\frac{1}{z}\left(\frac{b}{\sqrt{b^{2}+z^{2}}}-\frac{a}{\sqrt{a^{2}+z^{2}}}\right)\nonumber 
\end{align}

\subsubsection*{Discussion}

Our goal is to obtain the formula for the electric field of an infinite
wire. First, we cannot assume that a solution in the sense of (\ref{r2-limit-to-infinity})
exists. Therefore, we cannot set $a=-b$ and calculate the limit $b\rightarrow+\infty$
for $E_{x}$ and $E_{z}$ in (\ref{eq:wire-ExEz-integrals}). We cannot
assume only from symmetry that the field component parallel to the
wire, $E_{x}$, is zero as is usually done in approaches using Gauss'
law. Only after we prove that a solution exists -- which means that
we have to compute limits $a\rightarrow-\infty$ and $b\rightarrow+\infty$
independently -- then any symmetry-inspired methods or other shortcuts
can be used and would give the same result. These considerations may
seem superfluous, but such nuances play a crucial role in the case
of the infinite plate.

The results for $E_{x}$ and $E_{z}$ are independent of any order
in which limits $a\rightarrow-\infty$ and $b\rightarrow+\infty$
are calculated and in agreement with textbooks:

\begin{align*}
\lim_{b\rightarrow+\infty}\,\,\lim_{a\rightarrow-\infty}E_{x} & =0\\
\lim_{b\rightarrow+\infty}\,\,\lim_{a\rightarrow-\infty}E_{z} & =2k\lambda\frac{1}{z}
\end{align*}

\subsection{From the rectangle to the infinite plate}

In an analogy to the case of the finite wire, we start with a finite
rectangle and analyse what happens if sides of the rectangle are independently
extended to infinity. It will be shown that in some cases the integral
(\ref{r2-limit-to-infinity}) does not exist.

\subsubsection{The rectangle\label{sub:Rectangle}}

\begin{figure}
\noindent \begin{centering}
\includegraphics[width=0.8\textwidth]{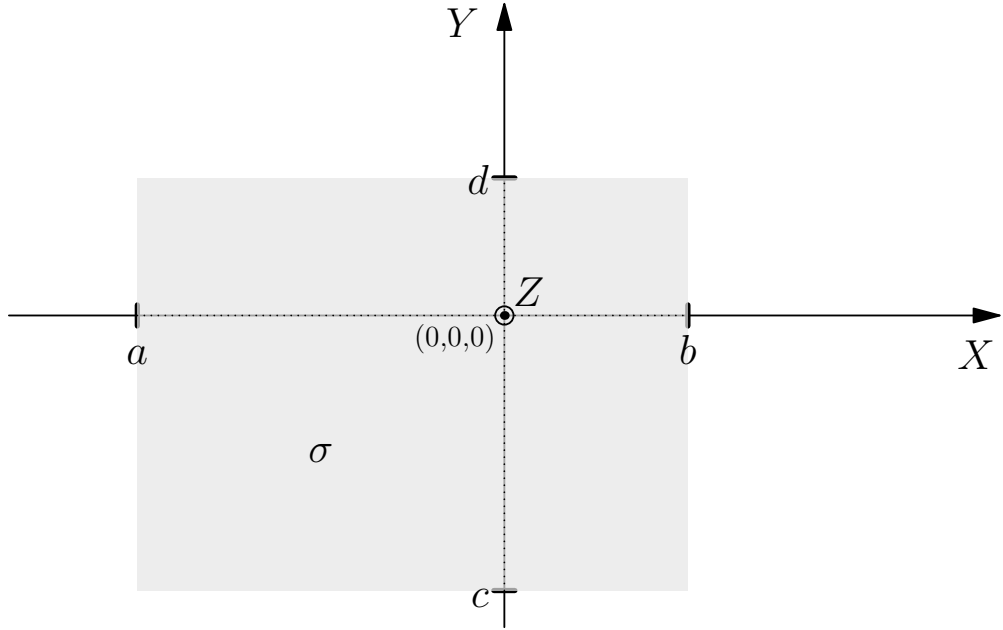} 
\par\end{centering}

\caption{\label{fig:Choice-of-coordinates}Choice of coordinates. The rectangle
$[a,\, b]\times[c,\, d]$ is charged with the constant surface density
$\sigma$. The electric field is determined on the $Z$ axis which
is perpendicular to the $XY$-plane.}
\end{figure}

Let us consider a two dimensional, uniformly charged rectangle $P=[a,\, b]\times[c,\, d]$
on the $XY$-plane. The choice of coordinates is shown in Fig.~\ref{fig:Choice-of-coordinates},
where $\sigma$ denotes a constant surface charge density. 

We determine the components of the electric field at point $\vec{r}=[0,\,0,\, z]$
on the $Z$ axis, assuming $z\neq0$. Details are presented in Appendix
\ref{Rectangle}. The $x$-component of the electric field is equal
to

\begin{equation}
E_{x}=k\sigma\ln\Biggl(\frac{d+\sqrt{b^{2}+d^{2}+z^{2}}}{c+\sqrt{b^{2}+c^{2}+z^{2}}}\,\frac{c+\sqrt{a^{2}+c^{2}+z^{2}}}{d+\sqrt{a^{2}+d^{2}+z^{2}}}\Biggr)\label{eq:Ex-1}
\end{equation}

As the result for $E_{y}$ can be easily obtained after a change of
variables in equation (\ref{eq:Ex-1})

\begin{equation}
E_{y}=k\sigma\ln\Biggl(\frac{b+\sqrt{d^{2}+b^{2}+z^{2}}}{a+\sqrt{d^{2}+a^{2}+z^{2}}}\,\frac{a+\sqrt{c^{2}+a^{2}+z^{2}}}{b+\sqrt{c^{2}+b^{2}+z^{2}}}\Biggr)\label{eq:Ey}
\end{equation}
we limit our considerations to $E_{x}$ only. The $z$-component of
the electric field is equal to 
\begin{eqnarray}
E_{z} & = & k\sigma\Biggl\{\arctan\Bigl[\frac{bd}{z\sqrt{b{}^{2}+d^{2}+z^{2}}}\Bigr]-\arctan\Bigl[\frac{bc}{z\sqrt{b{}^{2}+c^{2}+z^{2}}}\Bigr]\label{eq:Ez}\\
 &  & -\arctan\Bigl[\frac{ad}{z\sqrt{a{}^{2}+d^{2}+z^{2}}}\Bigr]+\arctan\Bigl[\frac{ac}{z\sqrt{a{}^{2}+c^{2}+z^{2}}}\Bigr]\Biggr\}\nonumber 
\end{eqnarray}

These results will be used in the following sections to calculate
the electric field of infinite charge distributions.

\subsubsection{From the rectangle to the infinite stripe}

We extend the rectangle to the infinite stripe by setting $d\rightarrow+\infty$
and $c\rightarrow-\infty$. A discussion about the limits would be
identical to the one from section \ref{sub:From-finite-to-Wire}.
After computing limits independently for $d$ and $c$ one obtains
(see Appendix \ref{infstripe}) well-defined components of the field

\begin{align}
E_{x\mathrm{\ stripe}} & =k\sigma\ln\frac{a^{2}+z^{2}}{b^{2}+z^{2}}\nonumber \\
E_{y\mathrm{\ stripe}} & =0\label{eq:E-Stripe}\\
E_{z\mathrm{\ stripe}} & =2k\sigma\Biggl\{\arctan\Bigl[\frac{b}{z}\Bigr]-\arctan\Bigl[\frac{a}{z}\Bigr]\Biggr\}\nonumber 
\end{align}

\subsubsection{From the stripe to the infinite plate\label{sub:From-the-stripe}}

This procedure breaks down if we ``extend'' the infinite straight
stripe to the infinite plate, calculating

\[
E_{x\mathrm{\ plane}}=\lim_{b\rightarrow+\infty}\,\,\lim_{a\rightarrow-\infty}E_{x\mathrm{\ stripe}}=\lim_{b\rightarrow+\infty}\,\,\lim_{a\rightarrow-\infty}k\sigma\ln\frac{a^{2}+z^{2}}{b^{2}+z^{2}}
\]

We aim to show that such a limit does not exist using a method similar
to the case of limit (\ref{sin}). To prove that various procedures
lead to different results, let us assume that 
\[
a=-\xi b
\]

where $\xi$ is an arbitrary constant, $\xi>0$. Then:

\begin{align}
E_{x\mathrm{\ plane}}=\lim_{b\rightarrow+\infty}\, E_{x\mathrm{\ stripe}} & =k\sigma\ln\xi^{2}\label{xplane}
\end{align}

It is clear that any result is obtainable. For example, if we set
$\xi=1$ then $E_{x\mathrm{\ plane}}=0$. But for $\xi=e$ one obtains
$E_{x\mathrm{\ plane}}=2k\sigma$. Similar reasoning shows that the
$y$-component also does not exist. To help students, we can use our
claims and explain the mathematical fact of non-existence of a limit
on the level of intuition: the electric field of the infinite plate
depends on the way the plate is built because different methods for
extending the stripe to infinity give different results. This means
that the electric field for the infinite plate does not exist. 

Problems with $E_{x\mathrm{\ plane}}$ and $E_{y\mathrm{\ plane}}$
do not influence the existence of the third limit for $E_{z\mathrm{\ plane}}$

\begin{eqnarray*}
E_{z\mathrm{\ plane}} & = & \frac{z}{|z|}\frac{\sigma}{2\varepsilon_{0}}
\end{eqnarray*}

The last result is presented in standard textbooks as the $z$ component
of the electric field of the infinite plate, the remaining components
are set to be zero as a result of \textquotedbl{}symmetry\textquotedbl{}.
However, with the help of formula (\ref{xplane}) we see that the
$E_{x\mathrm{\ plane}}$ and the $E_{y\mathrm{\ plane}}$ can be arbitrary
so we cannot talk about the vector quantity $\vec{E}_{\mathrm{\ plane}}$
in a meaningful way as two of its components are undefined.

\subsubsection{A quarter of $\mathbb{R}^{2}$\label{sub:Quarter-of-R^2}}

Another aspect of asymptotics of the electric field of the rectangle
from section \ref{sub:Rectangle} will be revealed if, instead of
extending opposite sides, one extends the rectangle to the first quarter
of the $XY$-plane by extending the adjacent sides. We set $a=0$,
$c=0$, $b\rightarrow+\infty$ and $d\rightarrow+\infty$. Then the
limit of the argument of the logarithm in equation (\ref{eq:Ex-1})
for $E_{x}$ equals zero:

\[
\lim_{b\rightarrow+\infty}\,\,\lim_{d\rightarrow+\infty}\Biggl(\frac{d+\sqrt{b^{2}+d^{2}+z^{2}}}{\sqrt{b^{2}+z^{2}}}\,\frac{\sqrt{z^{2}}}{d+\sqrt{d^{2}+z^{2}}}\Biggr)=0
\]

Thus one has

\[
E_{x\,\,{\mathbb{R}_{+}^{2}}}=\lim_{b\rightarrow+\infty}\,\,\lim_{d\rightarrow+\infty}E_{x}=-\infty
\]

One obtains the same result for $E_{y\,\,{\mathbb{R}_{+}^{2}}}$ in
equation (\ref{eq:Ey}) by calculating the same limit. Once more two
components of the electric field are undefined. The $z$-component
of the electric field is equal to

\begin{eqnarray*}
E_{z\,\,{\mathbb{R}_{+}^{2}}} & = & \frac{z}{|z|}\frac{\sigma}{8\varepsilon_{0}}
\end{eqnarray*}

which is a quarter of the standard solution for the $z$-component
of the field from an infinite plate. One could try to build the solution
for an infinite plate of four such quarters. Unfortunately, the vector
$\vec{E}_{\mathbb{R}_{+}^{2}}$ is undefined and the existence of
a well defined system made from four undefined subsystems cannot be
accepted in a mathematical and intuitive sense.

We showed that the solution for the infinite wire exists, but there
is no solution for the infinite plate. We did not find such a discussion
in any textbook. For example, in \cite{halliday2014instructors} (problem
33, p.~1014) students are encouraged only to calculate the field
of a half of an infinite wire. This result could be used to verify
the existence of a solution for the infinite wire. The next, natural
step would be to calculate the field from a half of an infinite plate.
That would necessarily lead to a discussion on the existence of a
solution.

\section{How important are finite size and asymmetry\label{sec:How-important-are}}

Although the problem of the existence of a solution for an infinite
plate is fundamental, it may be treated as the next academic curio.
A more practical question is: How much the field of a finite plate
differs from the widely used, standard textbooks values: $\frac{\sigma}{2\varepsilon_{0}}$
for a perpendicular component and zero for parallel component? The
results (\ref{eq:Ex-1}), (\ref{eq:Ey}), and (\ref{eq:Ez}) for a
uniformly charged rectangle can be used to analyze the ratios $E_{z}/\frac{\sigma}{2\varepsilon_{0}}$
and $E_{x}/E_{z}$. One expects the first ratio to be approximately
equal to 1, and the second to 0 if there is good agreement. As we
show in the following simple examples, for a wide range of parameters
values, the ratio $E_{z}/\frac{\sigma}{2\varepsilon_{0}}$ is around
$0.95$ as expected. However, the ratio $E_{x}/E_{z}$ can reach any
value. It is clear that the perpendicular field component cannot be
neglected, especially in calculations in which all components of the
electric field $\vec{E}$ are important.

\begin{figure}[!th]
\begin{centering}
\includegraphics[height=0.25\textheight]{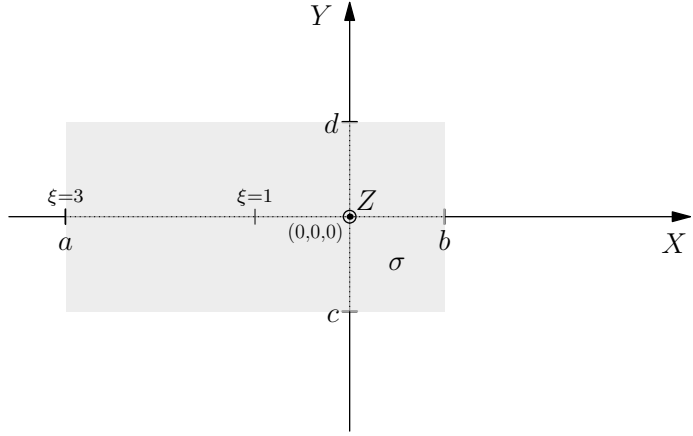} 
\par\end{centering}

\caption{\label{fig:coordinates-case-a}The extending stripe. The rectangle
$[a,\, b]\times[c,\, d]$ is charged with the constant surface density
$\sigma$. The $Z$ axis is perpendicular to the $XY$-plane. The
electric field is determined at point $\vec{r}=[0,\,0,\, z]$. Three
sides are set at $d=b=10z$ and $c=-10z$. The length of the rectangle,
and the position of the fourth side, is related to the asymmetry parameter
$\xi>0$ by setting $a=-10z\xi$. If $\xi=1$, the square is obtained.
As an example, the rectangle for the asymmetry parameter $\xi=3$
is shown.}
\end{figure}

\begin{figure}[!th]
\begin{centering}
\includegraphics[width=0.8\textwidth]{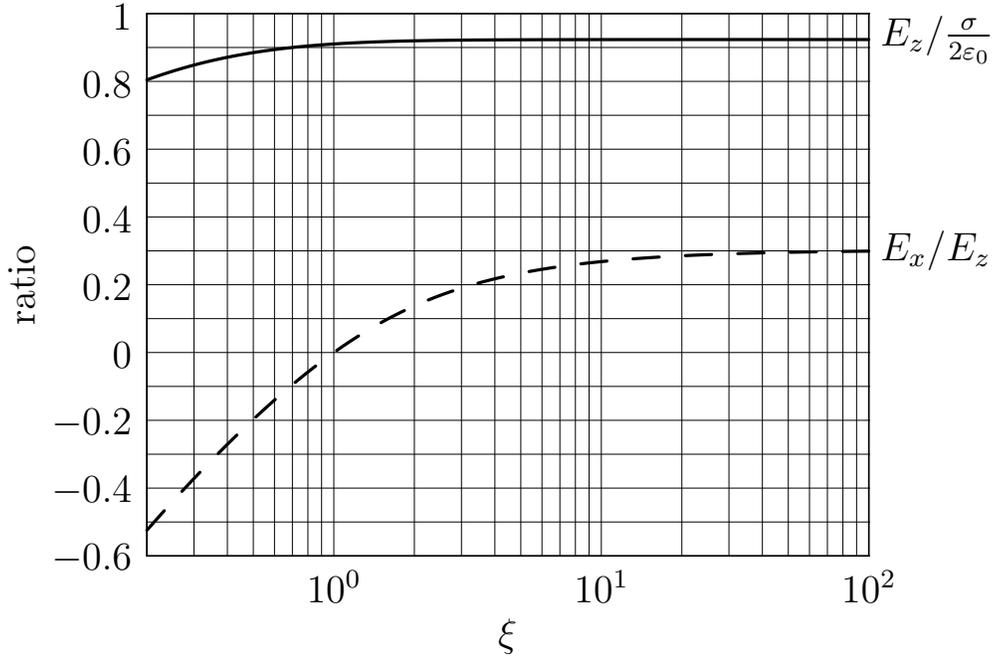} 
\par\end{centering}

\caption{\label{fig:ratios-case-a-1}The dependence of $E_{z}/\frac{\sigma}{2\varepsilon_{0}}$
and $E_{x}/E_{z}$ on the asymmetry parameter $\xi$ for the \emph{extending
stripe} case. The electric field is determined at point $\vec{r}=[0,\,0,\, z]$.
Three sides of the uniformly charged rectangle placed on $XY$-plane
are set at $d=b=10z$ and $c=-10z$ (Fig.~\ref{fig:coordinates-case-a}).
The length of the rectangle, and the position of the fourth side,
is related to the asymmetry parameter $\xi>0$ by setting $a=-10z\xi$.
For example, if $\xi=1$, the square is obtained (in this case $E_{x}$
is equal to $0$).}
\end{figure}

We demonstrate the behaviour of these ratios in two simple cases:

(a) \emph{An extending stripe. }The field is calculated in point $(0,\,0,\, z)$
where $z>0$ (Fig.~\ref{fig:coordinates-case-a}). To be in a reasonable
distance from the edges, we set the width of the rectangle to be 20
times larger than the distance $z$. Thus, we set three sides at $d=b=10z$
and $c=-10z$. The length of the rectangle, and the position of the
fourth side, we relate to the asymmetry parameter $\xi>0$ by setting
$a=-10z\xi$. For example, if $\xi=1$, the square is obtained. The
dependencies of $E_{z}/\frac{\sigma}{2\varepsilon_{0}}$ and $E_{x}/E_{z}$
on $\xi$ in this case are shown in Fig.~\ref{fig:ratios-case-a-1},
note that $E_{y}=0$. It is clear that $E_{x}$ cannot be neglected,
it is a significant component of the electric field: $|E_{x}/E_{z}|\gtrsim20\%$
for $\xi$ smaller than 0.5 or greater than 3.

It is worthwhile to comment about asymptotic behaviour: there are
finite, non-zero limits for $E_{z}$ and $E_{x}$ as $\xi\rightarrow\infty$
(this describes a stripe that is infinitely long on one side, here
-- on the negative part of the $X$ axis).

\begin{figure}[!th]
\begin{centering}
\includegraphics[height=0.25\textheight]{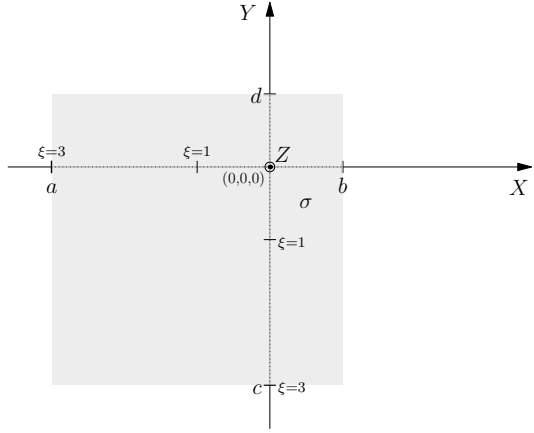} 
\par\end{centering}

\caption{\label{fig:coordinates-case-b}The extending square. The rectangle
$[a,\, b]\times[c,\, d]$ is charged with the constant surface density
$\sigma$. The $Z$ axis is perpendicular to the $XY$-plane. The
electric field is determined at point $\vec{r}=[0,\,0,\, z]$. Two
sides are set at $d=b=10z$. Both, the length and the width of the
rectangle are related to the asymmetry parameter $\xi>0$ by setting
$a=c=-10z\xi$. In this case, two sides of the resulting square ``move
away'' as the asymmetry parameter $\xi$ increases. As an example,
the square for the asymmetry parameter $\xi=3$ is shown.}
\end{figure}

\begin{figure}[!bh]
\begin{centering}
\includegraphics[width=0.75\textwidth]{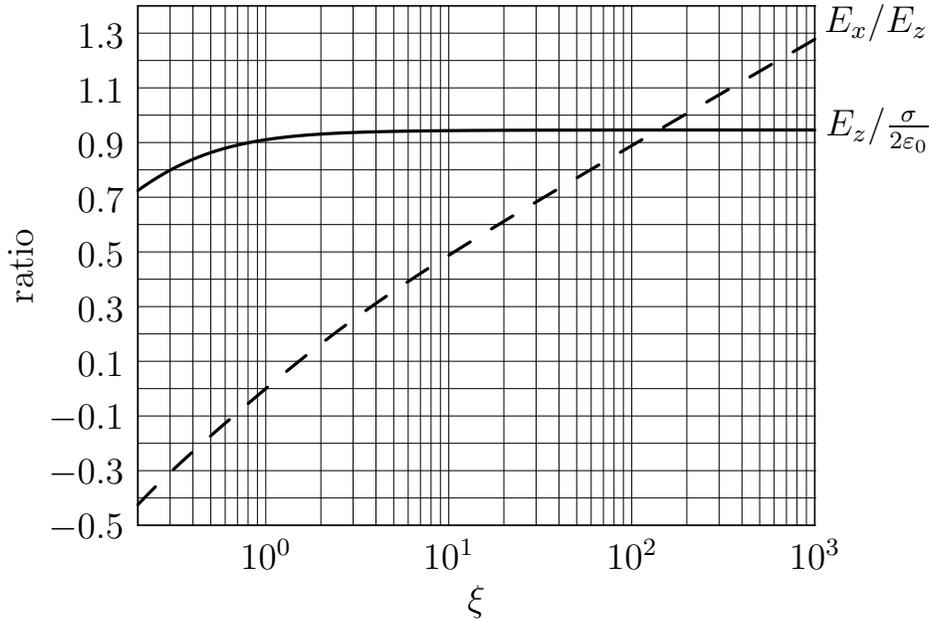} 
\par\end{centering}

\caption{\label{fig:ratios-case-b-1}The dependence of $E_{z}/\frac{\sigma}{2\varepsilon_{0}}$
and $E_{x}/E_{z}$ on $\xi$ for the \emph{extending square} case.
The electric field is determined at point $\vec{r}=[0,\,0,\, z]$.
Two sides of the uniformly charged rectangle placed on $XY$-plane
are set at $d=b=10z$ (Fig.~\ref{fig:coordinates-case-b}). Both,
the length and the width of the rectangle are related to the asymmetry
parameter $\xi>0$ by setting $a=c=-10z\xi$. In this case, two sides
of the resulting square ``move away'' as the asymmetry parameter
$\xi$ increases. For $\xi=1$ a center of the square is at point
$(0,\,0,\,0)$, and $E_{x}=0$ as expected.}
\end{figure}

(b) \emph{An extending square. }The field is calculated at point $(0,\,0,\, z)$
where $z>0$ (Fig.~\ref{fig:coordinates-case-b}). To be in a reasonable
distance from the edges we set the distance to the top and right edges,
of the rectangle to be 10 times the distance $z$ by setting $d=b=10z$.
Both, the length and the width of the rectangle we relate to the asymmetry
parameter $\xi>0$ by setting $a=c=-10z\xi$. In this case, two sides
of the resulting square ``move away'' as the asymmetry parameter
$\xi$ increases. The dependencies of $E_{z}/\frac{\sigma}{2\varepsilon_{0}}$
and $E_{x}/E_{z}$ on $\xi$ in this case are shown in Fig.~\ref{fig:ratios-case-b-1}.
It should be noted that $E_{y}=E_{x}$. For $\xi>2$ or $\xi<0.5$
the $E_{x}$ component of the field is greater than around 20\% of
$E_{z}$. For $\xi>200$ the $E_{x}$ component of the field is greater
than $E_{z}$. However, for $\xi>30$ the field component parallel
to the plate, $\sqrt{E_{x}^{2}+E_{y}^{2}}=\sqrt{2}|E_{x}$|, is greater
than $E_{z}$.

The asymptotic behaviour is different than in the case of \emph{the
extending stripe}. Only the perpendicular component, $E_{z}$, is
bounded. The parallel component is unbounded, $\lim_{\xi\rightarrow\infty}E_{x}=\infty$
and $\lim_{\xi\rightarrow\infty}E_{y}=\infty$, as in the case discussed
in section \ref{sub:Quarter-of-R^2}.

\begin{figure}[H]
\begin{centering}
\includegraphics[width=0.7\textwidth]{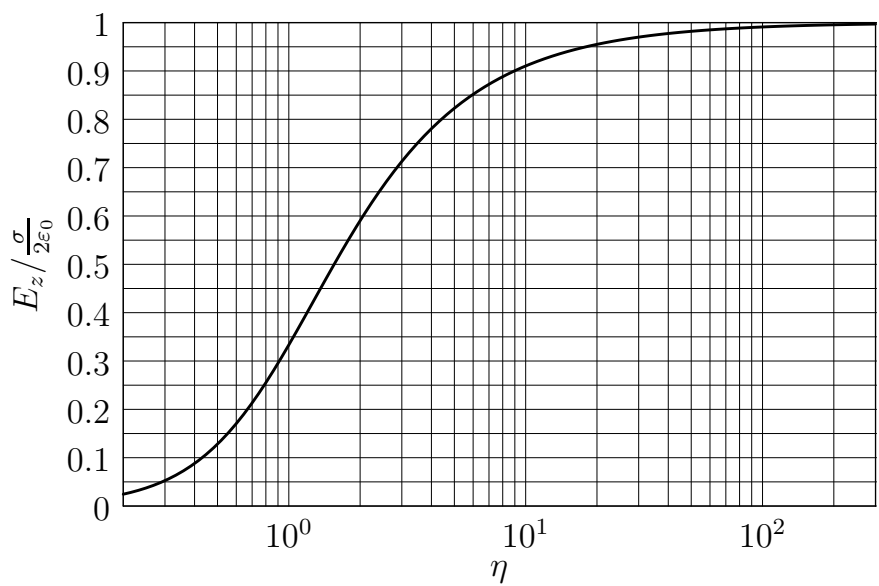} 
\par\end{centering}

\caption{\label{fig:ratio-Ez-only-case-c}The dependence of $E_{z}/\frac{\sigma}{2\varepsilon_{0}}$
above the centre of the extending square on the ratio $\eta$. The
electric field is determined at point $\vec{r}=[0,\,0,\, z]$. Four
sides of the uniformly charged rectangle placed on the $XY$-plane
are set at $b=d=\eta z$ and $a=c=-\eta z$ where $\eta>0$. Thus,
the length of a side of the square is equal to $2\eta z$. For example,
the field at the distance of $5$ cm above the centre of a square
with a side of length $60$ cm ($\eta=6$) would be equal to around
$85\%$ of $\frac{\sigma}{2\varepsilon_{0}}$.}
\end{figure}

For completeness we show how the field $E_{z}$ \emph{above the centre
of the extending square} varies\emph{. }The field is calculated at
point $(0,\,0,\, z)$ where $z>0$. To be above the centre of the
square we set $b=d=\eta z$ and $a=c=-\eta z$ where $\eta>0$. Thus,
the length of a side of the square is equal to $2\eta z$. The dependency
of $E_{z}/\frac{\sigma}{2\varepsilon_{0}}$ on the ratio $\eta$ in
this case, is shown in Fig.~\ref{fig:ratio-Ez-only-case-c}. It should
be noted that here $E_{y}=E_{x}=0$. If $\eta=1$, which means that
the length of a side of the square is equal to $2z$, the $z$-component
of the field is only around $35\%$ of $\frac{\sigma}{2\varepsilon_{0}}$.
The field magnitude reaches $95\%$ of $\frac{\sigma}{2\varepsilon_{0}}$
for $\eta=20$ (the length of a side of the square is equal to $40z$).
The field at the distance of $5$ cm above the centre of a square
with a side of length $60$ cm ($\eta=6$) would be equal to around
$85\%$ of $\frac{\sigma}{2\varepsilon_{0}}$.

\section{Conclusions}

We showed that for an infinite, uniformly charged plate no well defined
electric field exists in the framework of electrostatics. We propose
heuristic tools (the claims) that would help to align intuitions in
the spirit of the rigorous definition of an integral. We want students
to first consider the existence of the solution. We demonstrated that
unfortunately some classical problems present in textbooks cannot
be defined in a meaningful way -- it is hard to talk about an electric
field when only one component of the vector quantity is not ill-defined.
Such problems seem to be very simple but their simplicity is deceptive.

The good news is that a discussion about the applicability of solutions
for a finite plate to an \textquotedbl{}infinite plate\textquotedbl{}
problem is relatively simple. The transition from a rectangle to an
infinite plate can lead through an infinite stripe or a quarter of
$\mathbb{R}^{2}$ and help to understand where the solution ceases
to exist. As we showed, a more rigorous discussion during classes
is possible. Moreover, it may be interesting for students as a working
example of the advantages of taking a closer look at definitions of
mathematical objects. The didactic challenge can be overcome.

\vspace{2em}
 {The authors would like to thank Kazimierz Napiórkowski, Andrzej
Majhofer and~Robin \& Tad Krauze for fruitful discussions and valuable
comments.} \bibliographystyle{plain} 

\appendix

\section{Uniformly charged rectangle}

\label{Rectangle}

Let us consider a two dimensional, uniformly charged -- with constant
surface charge density $\sigma$ -- rectangle $P=[a,\, b]\times[c,\, d]$
on the $XY$-plane. The choice of coordinates is shown in Fig.~\ref{fig:Choice-of-coordinates}.
We determine the electric field at point $\vec{r}=[0,\,0,\, z]$ on
the $Z$ axis, assuming $z\neq0$. Using Coulomb's law and superposing
contributions from infinitesimal charge elements $\sigma\,\mbox{d}S'$
at point $\vec{r}'=[x',\, y',\,0]$ one obtains:

\begin{eqnarray*}
\vec{E}(\vec{r}) & = & k\sigma\int_{P}\frac{\vec{r}-\vec{r}'}{|\vec{r}-\vec{r}'|^{3}}\mbox{d}S'
\end{eqnarray*}

where

\[
\int_{P}\frac{\vec{r}-\vec{r}'}{|\vec{r}-\vec{r}'|^{3}}\mbox{d}S'=\hat{e}_{x}\int_{P}\frac{-x'}{|\vec{r}-\vec{r}'|^{3}}\mbox{d}S'+\hat{e}_{y}\int_{P}\frac{-y'}{|\vec{r}-\vec{r}'|^{3}}\mbox{d}S'+\hat{e}_{z}\int_{P}\frac{z}{|\vec{r}-\vec{r}'|^{3}}\mbox{d}S'
\]

and

\begin{eqnarray*}
\vec{r}-\vec{r}' & = & [-x',\,-y',\, z]\\
|\vec{r}-\vec{r}'| & = & \sqrt{x'^{2}+y'^{2}+z^{2}}
\end{eqnarray*}

$x'\in[a,\, b]$ and $y'\in[c,\, d]$ at $z'=0$. Let us focus on
the $x$-component of $\vec{E}$:

\[
E_{x}=k\sigma\int_{c}^{d}\int_{a}^{b}\frac{-x'}{\sqrt{x'^{2}+y'^{2}+z^{2}}^{3}}\mbox{d}x'\,\mbox{d}y'
\]

After the first integration one obtains

\begin{eqnarray*}
\int_{a}^{b}\frac{-x'}{\sqrt{(x'^{2}+y'^{2}+z^{2})^{3}}}\mbox{d}x' & = & \left.\frac{1}{\sqrt{x'^{2}+y'^{2}+z^{2}}}\right|_{a}^{b}\\
 & = & \frac{1}{\sqrt{b^{2}+y'^{2}+z^{2}}}-\frac{1}{\sqrt{a^{2}+y'^{2}+z^{2}}}
\end{eqnarray*}

The second integration leads to

\begin{eqnarray*}
\int_{c}^{d}\frac{1}{\sqrt{b^{2}+y'^{2}+z^{2}}}\mbox{d}y' & = & \left.\ln|y'+\sqrt{b^{2}+y'^{2}+z^{2}}|\right|_{c}^{d}\\
 & = & \ln\left|\frac{d+\sqrt{b^{2}+d^{2}+z^{2}}}{c+\sqrt{b^{2}+c^{2}+z^{2}}}\right|
\end{eqnarray*}

Finally, the $x$-component of the electric field is equal to:

\begin{equation}
E_{x}=k\sigma\ln\Biggl(\frac{d+\sqrt{b^{2}+d^{2}+z^{2}}}{c+\sqrt{b^{2}+c^{2}+z^{2}}}\,\frac{c+\sqrt{a^{2}+c^{2}+z^{2}}}{d+\sqrt{a^{2}+d^{2}+z^{2}}}\Biggr)\label{eq:Ex}
\end{equation}

The result for $E_{y}$ can be easily obtained after a change of variables
in equation (\ref{eq:Ex}).

To fully describe the electric field of the uniformly charged rectangle
we calculate the $z$ component of $\vec{E}$:

\[
E_{z}=k\sigma\int_{c}^{d}\int_{a}^{b}\frac{z}{\sqrt{x'^{2}+y'^{2}+z^{2}}^{3}}\mbox{d}x'\,\mbox{d}y'
\]

The first integration:

\begin{eqnarray*}
\int_{a}^{b}\frac{1}{\sqrt{x'^{2}+y'^{2}+z^{2}}^{3}}\mbox{d}x' & = & \frac{x'}{(y'^{2}+z^{2})\sqrt{x'^{2}+y'^{2}+z^{2}}}\Bigr|_{a}^{b}\\
 & = & \frac{b}{(y'^{2}+z^{2})\sqrt{b{}^{2}+y'^{2}+z^{2}}}-\frac{a}{(y'^{2}+z^{2})\sqrt{a{}^{2}+y'^{2}+z^{2}}}
\end{eqnarray*}

The next integral is more complicated: 
\begin{eqnarray*}
\int_{c}^{d}\frac{b}{(y'^{2}+z^{2})\sqrt{b{}^{2}+y'^{2}+z^{2}}}\,\mbox{d}y' & = & \frac{1}{z}\arctan\Bigl[\frac{by'}{z\sqrt{b{}^{2}+y'^{2}+z^{2}}}\Bigr]\Bigr|_{c}^{d}\\
 & = & \frac{1}{z}\Bigl\{\arctan\Bigl[\frac{bd}{z\sqrt{b{}^{2}+d^{2}+z^{2}}}\Bigr]\\
 &  & -\arctan\Bigl[\frac{bc}{z\sqrt{b{}^{2}+c^{2}+z^{2}}}\Bigr]\Bigr\}
\end{eqnarray*}

Finally, we obtain

\begin{eqnarray*}
E_{z} & = & k\sigma\Biggl\{\arctan\Bigl[\frac{bd}{z\sqrt{b{}^{2}+d^{2}+z^{2}}}\Bigr]-\arctan\Bigl[\frac{bc}{z\sqrt{b{}^{2}+c^{2}+z^{2}}}\Bigr]\\
 &  & -\arctan\Bigl[\frac{ad}{z\sqrt{a{}^{2}+d^{2}+z^{2}}}\Bigr]+\arctan\Bigl[\frac{ac}{z\sqrt{a{}^{2}+c^{2}+z^{2}}}\Bigr]\Biggr\}
\end{eqnarray*}

It is simple to show that

\begin{eqnarray*}
\lim_{b\rightarrow+\infty}\,\,\lim_{a\rightarrow-\infty}\lim_{d\rightarrow+\infty}\,\,\lim_{c\rightarrow-\infty}E_{z} & = & \frac{z}{|z|}k\sigma\Biggl\{\frac{\pi}{2}+\frac{\pi}{2}+\frac{\pi}{2}+\frac{\pi}{2}\Biggr\}=\frac{z}{|z|}\frac{\sigma}{2\varepsilon_{0}}
\end{eqnarray*}

This limit for $E_{z}$ is equal to the result well known from textbooks.

\section{From a rectangle to an infinite stripe\label{infstripe}}

We ``extend'' the rectangle to the infinite stripe by setting $d\rightarrow+\infty$
and $c\rightarrow-\infty$:

\begin{eqnarray*}
\lim_{c\rightarrow-\infty}\,\,\lim_{d\rightarrow+\infty}\Biggl(\frac{d+\sqrt{b^{2}+d^{2}+z^{2}}}{c+\sqrt{b^{2}+c^{2}+z^{2}}}\,\frac{c+\sqrt{a^{2}+c^{2}+z^{2}}}{d+\sqrt{a^{2}+d^{2}+z^{2}}}\Biggr) & = & \lim_{c\rightarrow-\infty}\Biggl(\frac{c+\sqrt{a^{2}+c^{2}+z^{2}}}{c+\sqrt{b^{2}+c^{2}+z^{2}}}\Biggr)\\
 & = & \frac{a^{2}+z^{2}}{b^{2}+z^{2}}
\end{eqnarray*}

One obtains a well defined $x$-component of the field: 
\[
E_{x\mathrm{\ stripe}}=k\sigma\ln\frac{a^{2}+z^{2}}{b^{2}+z^{2}}
\]

\section{Examples of inconsistencies\label{sec:Examples-of-inconsistencies}}

We are aware that it is a risky task to pinpoint the inconsistencies
in well established textbooks. However, as university teachers that
have to explain the issue to confused students every year, we would
be more than satisfied to be able to recommend a textbook in which
the authors present a consistent approach to problems with infinite
charge distributions. Unfortunately, we did not find a mathematically
correct treatment of such cases. To show that the problem is widespread,
we present an arbitrary list of a few introductory courses in electrostatics
in which the existence of the electric field or the force due to an
uniformly charged infinite object is taken for granted. 
\begin{itemize}
\item In \cite{halliday2014fundamentals} (\emph{Cancelling Components},
pp.~639-640) the authors explain that in the case of a uniformly
charged ring the components perpendicular to the ring axis are cancelled.
This result is used as well in the case of a uniformly charged disk
(pp.~643-644). However, at the end of this section the authors obtain
the electric field for an infinite plate by extending the radius of
the disk to infinity. There is no discussion of the existence of the
presented integral if the radius of the disk is infinite. So, the
components perpendicular to the axis of the disk are obtained on the
same basis as in result (\ref{int_sin_minus_B_to_B}). In the following
(p.~673) or in \cite{Dobbs} (p.~13), the field from an infinite
sheet is calculated using Gauss' law, with the same assumption that
the field parallel to the plate is zero. As we show in section \ref{sub:From-the-stripe}
or \ref{sub:Quarter-of-R^2}, this field does not exist in the framework
of electrostatics. 
\item In \cite{Prytz2015} (p.~51) the infinite sheet is built from infinite
wires. The author observes that the integrand is an odd function,
thus the result must be zero. Once more, students may think about
$\sin(x)$ as the integrand (see Eq.~\ref{int_sin_minus_inf_to_inf})
and wonder why physics lectures are not compatible with mathematical
ones. 
\item We find in \cite{Feynman2013} (section 13-4, pp.~from 13-13 to 13-14)
that in the case of an infinite plate only the perpendicular component
of the gravitational or the electric field is considered. 
\item In \cite{halliday2014instructors} (problem 33, p.~1052) and in \cite{MIT}
(section 4.8, p.~31) we have examples of the standard superposition
of the fields from infinite plates. It is similar to superposing undefined
quantities -- such are the field components parallel to an infinite
sheet. 
\item In \cite{halliday2014instructors} (problem 37, p.~1053) the authors
instruct students on how they should think: \emph{''THINK To calculate
the electric field at a point very close to the center of a large,
uniformly charged conducting plate, we replace the finite plate with
an infinite plate having the same charge density. Planar symmetry
then allows us to apply Gauss' law to calculate the electric field.''
} 
\item In \cite{Griffiths2013electrodynamics} (p.~53) we read \emph{``In
some textbook problems the charge itself extends to infinity (we speak,
for instance, of the electric field of an infinite plane, or the magnetic
field of an infinite wire). In such cases the normal boundary conditions
do not apply, and one must invoke symmetry arguments to determine
the fields uniquely.'' } This suggests that the authors do not doubt
that electrostatics is able to describe the case. The only problem
is how to change the game rules to prove the result we believe in. 
\item In \cite{herbert1991introductory} (pp.~45-46) the field components
parallel to an infinite sheet are calculated and zero values are obtained.
The author integrates first over an azimuthal angle, as the result
is zero, the next integration over a radius is not necessary. However,
students who usually already know Fubini's theorem can try to integrate
first over the radius that leads them to infinity! 
\end{itemize}
In all these cases the existence of the solution is assumed, and the
authors' main goal is to obtain a mathematical formula, usually via
some technical shortcut. A discussion of the existence of the solution
would be beneficial for the didactic process, and is likely to lead
to the correct result. 
\end{document}